\documentstyle[prd,aps,preprint]{revtex}
\tightenlines
\input epsf.tex
\def\DESepsf(#1 width #2){\epsfxsize=#2 \epsfbox{#1}}

\newcommand{\be}{\begin{equation}}
\newcommand{\ee}{\end{equation}}
\newcommand{\bea}{\begin{eqnarray}}
\newcommand{\beas}{\begin{eqnarray*}}
\newcommand{\eea}{\end{eqnarray}}
\newcommand{\eeas}{\end{eqnarray*}} 
\newcommand{\ba}{\begin{array}}
\newcommand{\ea}{\end{array}}

\begin{document}

\draft
\preprint{\vbox{
\hbox{UMD-PP-00-033}}}

\title{Sterile Neutrino as a Bulk Neutrino}

\author{ R. N. Mohapatra$^1$\footnote{e-mail:rmohapat@physics.umd.edu},
and 
A. P\'erez-Lorenzana$^{1,2}$\footnote{e-mail:aplorenz@Glue.umd.edu} }

\address{
$^1$ Department of
Physics, University of Maryland, College Park, MD, 20742, USA\\
$^2$  Departamento de F\'\i sica, 
Centro de Investigaci\'on y de Estudios Avanzados del I.P.N.\\
Apdo. Post. 14-740, 07000, M\'exico, D.F., M\'exico.}
\date{October, 1999}
\maketitle
\begin{abstract}
{If light sterile neutrinos are needed to understand the neutrino
puzzles, as is currently indicated, a major theoretical challenge is to
understand
why its mass is so small. It is a more serious problem than understanding
the small mass of the familiar neutrinos. We discuss a new way to
solve this problem by identifying the sterile neutrino as gauge neutral
fermion propagating in the bulk of a higher dimensional theory, with bulk
size of order of a millimeter. The smallness of its mass is then a
consequence of the size of the extra dimension and does not need the
introduction of new symmetries. We present a realistic model for neutrino
masses and mixings that implements this idea.}
\end{abstract} 

\vskip0.5in

\section{Introduction}

Recent experimental data on atmospheric neutrinos from the
Super-Kamiokande~\cite{sk}
collaboration has provided conclusive evidence for neutrino oscillations
and confirms earlier indications of such oscillations from other
experiments~\cite{atmos}. In addition there is also evidence for neutrino
oscillations from the solar neutrino  deficit observed by Kamiokande,
Homestake, Gallex, Sage and Super-Kamiokande~\cite{expt,superK}, and   
direct observation of $\nu_\mu \rightarrow \nu_e$
($\bar\nu_\mu \rightarrow \bar\nu_e$) in the LSND~\cite{LSND} experiment.
All these suggest nonvanishing
masses for at least two of the three standard neutrinos and
opens the window to explore physics beyond the standard model. 

It has been realized that to explain all three evidences for neutrino
oscillations three different mass differences are needed, while only two
can be obtained with three neutrinos. Solar neutrino  data
requires~\cite{fogli}
 \bea
 \Delta m^2_{sol}&\sim& 3\times 10^{-6} - 1.2 \times 10^{-5}~ eV^2,
 \nonumber\\
 \sin^22\theta_{sol} &\sim& 3\times 10^{-3} - 1.5 \times 10^{-2}
 \label{sol}
 \eea
for the small mixing angle MSW~\cite{MSW} solution. 
Atmospheric neutrinos seems to prefer $\nu_\mu - \nu_\tau$
oscillations  with a maximal mixing~\cite{gg}:
 \bea
 \Delta m^2_{atm}&\sim& 4\times 10^{-4} - 5 \times 10^{-3}~ eV^2,
 \nonumber\\
 \sin^22\theta_{atm} &\sim& .76 - 1.
  \label{atm}
 \eea
Finally, the LSND results along with other constraints from
KARMEN~\cite{karmen}, 
Bugey~\cite{bugey} and  E776 at BNL~\cite{E776} suggest
 \bea
 \Delta m^2_{e\mu}&\sim& .2 - 2  ~ eV^2,
 \nonumber\\
 \sin^22\theta_{e\mu} &\sim& 2\times 10^{-3} - 4 \times 10^{-2}.
 \label{emu}
 \eea

A simple explanation of these parameters could be obtained if there is a
sterile neutrino~\cite{sterile} (which is a light neutrino that does not
couple to the Standard Model particles). The existence of a light
sterile neutrino however poses a major theoretical challenge. It is more
problematic to understand its small mass than to understand the small mass
of the familiar neutrinos. This is because, the familiar neutrinos are
standard model gauge nonsinglets; therefore their mass can at most be of
order of the weak scale. On the other hand the sterile neutrinos being
standard model singlets, their apriori mass could be of order of the
Planck scale- some sixteen orders of magnitude higher.
 
Several scenarios involving sterile neutrino have been studied in the
literature~\cite{sterile2,foot,smirnov,sterile3} where the general
strategy has been to invoke new symmetries of different kinds to explain
its small mass. It is clearly too early to say which if any of these
approaches is the correct one. Here we explore a completely different
approach to its mass. We consider the possibility that the sterile
neutrino is a massless Dirac bulk fermion in a higher dimensional
brane-bulk picture with
at least one large extra dimension. The bulk fermion has Kaluza-Klein (KK)
excitations whose spacing is given by the inverse size of the bulk. If the
bulk size is large (say a millimeter) as has been speculated
recently\cite{many3}, then the sterile neutrino (i.e. the bulk neutrino)
mass is automatically small without the need for any extra symmetries.
In recent days,
interest in such theories (i.e. theories with large extra dimensions) has
been heightened from independent experimental considerations since they
lead to deviations from Newton's inverse square law of gravity at small
distances. In fact the maximum bulk size believed to be at the boundary of
gravity experiments is a millimeter, which corresponds to sterile neutrino
masses of milli-eV's. This value of masses is precisely what is
interesting for solving the solar neutrino problem providing an intriguing
connection between the neutrino experiments and those searching for
deviations from Newton's law in the millimeter range.

In the brane-bulk models with large extra dimensions, the smallness of the
KK excitations of the bulk fermion mass is related to the size of the
extra dimensions as just explained. On the other hand the small mass of
familiar neutrinos arises from the following fact. There is a
suppression of the coupling between
neutrinos living on the bulk with those fixed on our four dimensional
world~\cite{dienes,dvali,pospel,mnp} which arises naturally from
normalization of the bulk field. As far as the oscillations between the
brane and bulk neutrinos is concerned, it is dominated
by the mass of the lowest KK mode of the
bulk neutrino. Thus one has the right parameter range to give an
explanation for the solar neutrino deficit.

Two distinctive scenarios which involve bulk neutrinos have been outlined
in recent literature. The first uses only the standard model left handed
neutrinos in the brane and one or more bulk fermions\cite{dienes,dvali}
and a second one that has both the left and right handed neutrinos in the
brane in combination with a bulk neutrino\cite{mnp}. In the second case,
one could invoke the seesaw mechanism~\cite{seesaw} 
in the brane to understand the
smallness of the neutrino masses; therefore the higher dimensional physics
need not play a role. Of course it could happen that implementation of 
seesaw mechanism
is impossible due to group theory of the model\cite{mnp} in
which case the extra dimension as well as the string scale will play a
role in neutrino physics. 

It must however be emphasized that even though
mechanisms to understand small neutrino masses have been outlined, no
attempt has been made to construct realistic models that explain neutrino
observations using the property of higher dimensional physics. It is the
goal of this paper to make an attempt in this direction and in particular
construct a model involving a sterile neutrino that explains the present
neutrino data.

We first study the models without right handed neutrinos and point out
that even though the mass of the sterile neutrino is naturally small in
these models, several unnatural assumptions are needed if we are to obtain
a desirable mixing and mass pattern for understanding observations.
We then study models with both left and right handed neutrinos in the
brane with seesaw mechanism that explains the small masses for two of the
known
neutrinos and include a bulk neutrino with one extra dimension being
in the submillimeter range to play the role of the sterile neutrino. Since 
the masses of many of the low KK excitations are then in the range
of interest for solar neutrino oscillations, the key question is
whether the mixings pattern is such as to be of interest in solving the
neutrino puzzles without at the same time contradicting known
observations. In particular, we explore
whether only the lowest mode of this bulk neutrino can play the role of
the sterile neutrino. The advantage of this approach over other models for
the sterile neutrino is that smallness of its mass is now a
geometrical rather than a symmetry effect. The nontriviality of the
problem arises from the fact that the bulk neutrino of course has infinite
number of excitations and one has to tackle the mathematical problem of
extracting the physical masses and mixings of neutrinos from this complex
situation and study whether one has a solution to the neutrino puzzles.

Our basic results in the second case are the following: the lightest
eigenstate which is predominantly the
electron neutrino is massless and it mixes with the first excited KK mode
of the bulk neutrino, which has mass of order $R^{-1}$. We use this to 
solve the
solar neutrino deficit via the small angle MSW mechanism. This requires
the size of
the extra dimension to be in the micro to millimeter range and a
string scale which is in the range of $10^{8}$ GeV. Only one large extra
dimension is
sufficient in our discussion. The smallness of the $\nu_{\mu,\tau}$ is due
to seesaw mechanism in the brane that uses intermediate
seesaw scales (of the order of the string scale or so). The maximal
$\nu_\mu-\nu_\tau$ mixing needed for atmospheric oscillations owes its
origin to the texture of the right handed neutrino matrix. 
The LSND data is
explained in a natural way via small mixing between generations. Thus
the new contribution of this paper is the identification of the
lowest KK mode of the bulk neutrino as a viable sterile neutrino
candidate and its embedding into a realistic four neutrino gauge
theory framework.

\section{Models without right handed neutrinos}

Let us start by  summarizing the simplest mechanism to produce neutrino
masses using extra dimensions as discussed in Ref.~\cite{dvali}. One 
assumes that all the Standard Model particles are localized on a brane
embedded in the bulk of larger dimensions. The 
conservation of gauge flux then implies that, besides gravity, the only
fields that could propagate in the extra dimensions are standard model
gauge singlets. Their coupling with
the brane fields is naturally suppressed by the volume factor $M_*\over
M_{P\ell}$, where, from the observed strength  of gravity the Plank scale
and the string scale $M_*$ are related by
 \be
 M^2_{P\ell}={ M_{*}}^{n+2}V_n
 \label{mpl}
 \ee
with $V_n$ the volume of the extra space. We will assume  that one of the
radii of the compact
extra dimensions, R, is larger.  Now, by including the
coupling of a bulk neutrino $\nu_B( x^{\mu}, y)$ to the standard model
lepton doublet $L(x^{\mu}, y=0)$ we get the following terms which are
responsible for the neutrino mass in this model
 \be
 {\cal S} = \kappa \int d^4x \bar{L} H \nu_{BR}(x, y=0) + \int d^4x dy
 \bar{\nu}_{BL}(x,y)\partial_5 \nu_{BR}(x,y) + h.c.
 \ee
where $H$ is the Higgs doublet, $L$ is the lepton doublet and $\kappa$ is
the suppressed Yukawa coupling
 \be
 \kappa = h {M_*\over M_{P\ell}}.
 \label{kappa}
 \ee
By introducing the Fourier expansion of the bulk field  we may write
down the Dirac mass terms in  (\ref{mpl})  as 
 \be 
  (\bar{\nu}_{eL}  \bar{\nu}'_{BL})\left(\begin{array}{cc}
 m &\sqrt{2} m\\ 0 & \partial_5
 \end{array}\right)\left(\begin{array}{c}\nu_{0B} \\
 \nu'_{BR}\end{array}\right)
 \label{m1}
 \ee
where $\nu'_B$ embodies in a compact way the KK excitations along the fifth
dimension and $m= \kappa v$ is the mass term  produced by 
the vacuum expectation value, $v$,  of the Higgs field.
The entry $\sqrt{2} m$  has to be interpreted as an (infinite) row vector
of the
form $\sqrt{2} m (1,1,\cdots)$. 
The operator $\partial_5$ represents the diagonal and infinite mass
matrix of the KK modes.
The eigenvalues of this operator are $n\mu_0$,
with $\mu_0=1/R$. For $m \ll \mu_0$
the mixing angle of the standard neutrino with the
n-th bulk mode is~\cite{dvali}
 \be
 \tan\theta_n\approx {\xi \over n};
 \qquad \mbox{with}\qquad \xi = \sqrt{2}m R .
 \label{tan1}
 \ee
Then, the standard neutrino has a mass $m$ and oscillations into
bulk neutrino are present dominated by the first exited mode of mass 
$\mu_0$. 
An explanation to the solar neutrino problem then follows  if
one assumes MSW mechanism with $\mu_0 \simeq 10^{-3}$ eV ($R\simeq 0.2.$
mm). The question now is how one can extend the model to incorporate the
atmospheric data and the LSND results.

First thing to note is that in this scenario, the mixing between neutrinos
of different generations is expected to be produced through the couplings
with the bulk neutrino only. However, even if the bulk neutrino is not
blind
and couples with a different Yukawa coupling to each brane neutrino, the
mass matrix will produce two massless neutrinos which decouple from the
oscillation pattern. Therefore, there is no room for understanding the
atmospheric and LSND data in this framework.

Let us now see if this problem can be cured by including three bulk
neutrinos. The most general Dirac  mass in this case may be written,
after a proper rotation of the bulk fields, as
 \be 
 {\cal L} = \bar{\bf \nu}_L\cdot V M_D\cdot {\bf \nu}_{BR} (y=0)
 + \int dy\, \bar{\bf \nu}_{BL}\cdot\partial_5 \cdot{\bf \nu}_{BR}
 + h.c.
 \label{3bulk} 
\ee
with ${\bf \nu}_L= (\nu_e,\nu_\mu,\nu_\tau)_L$; 
${\bf \nu}_B= (\nu^1_B,\nu^2_B,\nu^3_B)$; $M_D= Diag(m_1,m_2,m_3)$;
and where the unitary matrix $V$ depends on the texture of the Yukawa
couplings to bulk neutrinos. The mass parameters in $M_D$ are small
numbers  of the order of eV or so and are given by the eigenvalues of the
Yukawa matrix multiplied by $v$. Since the  extra dimensional volume 
suppression is present, it is not unnatural to choose one of the masses 
to be  in the range of $10^{-4}$ eV as in (\ref{kappa}) to solve the solar
neutrino puzzle.

After rotating  the standard sector  by $V$, the last 
expression simplifies to the form
 \be 
 {\cal L} =
 \sum_{a=1,2,3}\left[ m_a \bar\nu^a_L \nu^a_{BR}(y=0) + 
 \int dy\, \bar\nu^a_{BL}\partial_5 \nu^a_{BR} 
 + h.c. \right]
\label{three} 
 \ee
Therefore, in this picture the pattern of neutrino oscillations follows
the same as before with a generational mixing given by $V$, which must 
provide atmospheric and LSND mixings. However a proper understanding of
the neutrino puzzles would need
three independent mass parameters. At first glance we would be tempted to
believe that these masses could be those involved in  (\ref{three}).
However, this is not the case. As we shall show below the heaviest
eigenstate $\nu^a_L$ has a mass  $\tilde m_a=\min \{ \mu_0/2, m_a \}$.
Thus, if solar neutrino data is assumed to be solved by $\nu_e-\nu_B$
oscillation, this would require that $\mu_1\simeq 10^{-3}$ eV. As a result 
other mass differences in this picture become
too small to solve the atmospheric and LSND data. If on the other hand, we
assumed that it is the atmospheric neutrino puzzle which is solved by
using the bulk neutrino, then we would choose $\mu_1\simeq 0.06$ eV; one
could envision solving the solar neutrino puzzle by $\nu_e-\nu_{\mu}$
oscillation, with the relevant mass difference coming from small Yukawa
couplings. Then we would be unable to explain the LSND data.

Thus it appears that the only way out is to
assume that the bulk neutrinos come from different branes 
with different sizes:
$\mu_0\simeq 10^{-3}$, and $\mu_{2,3}\simeq 1$ eV, with 
$\Delta m^2_{atm} = \mu_2 - \mu_3$. The challenge in such a
scenario is to
explain how the mass terms  can be diagonalized simultaneously, a fact
used to obtain Eq. (\ref{three}). 

To prove our previous statement, let us consider again the
mass matrix (\ref{m1}).
In this compact notation it is simple to get the exact solution for the
eigensystem even when the mass matrix is infinite. 
First the characteristic equation is given by
 \be 
 (m_n^2 - \partial_5^2)\left[m_n^2 - m^2 + 
 {2 m_n^2 m^2 \over  \partial_5^2 - m_n^2}\right] = 0 ,
 \ee
where $m_n$ is the mass eigenvalue and a sum  on the last term is
implicit. This expression 
translates into the same result that was obtained
before in references~\cite{dienes,dvali}
 \be 
 m_n = \pi m^2 R \ \cot(\pi m_n R).
 \label{char1}
 \ee
The  eigenstates are thus given symbolically by 
 \be 
 \tilde \nu_{nL} = {1\over N_n} \left[ \nu_L + 
 {\sqrt{2} m \partial_5\over m_n^2 - \partial_5^2 }\, \nu'_{BL}\right],
 \ee
where the normalization factor $N_n$ is easily computed to be
 \be
 N^2_n = 1 + 
 2 m^2 R^2 \sum_{k=1}^\infty{k^2\over (k^2 - m_n^2R^2)^2} = 
 {1\over 2} \left[  1 + (\pi m R)^2  + \left({m_n\over m}\right)^2 \right].
 \ee
 Now we may express $\nu_L$ in terms of the massive modes just as 
 \be
 \nu_L = \sum_{n=0}^\infty {1\over N_n} \tilde \nu_{nL}.
 \ee
Since $N_n$ has a minimum for the lowest value of $m_n$, the main
component of $\nu_L$ is always the lightest mode $\tilde\nu_{0L}$.  
Notice from (\ref{char1}) 
that when $mR\ll 1$, the eigenvalues comes out to be
$m_0=m$ and $m_n = n\mu_0$ for $n>0$, and  $N_n\approx m_n/\sqrt{2} m$, 
thus, we recover the mixing angle  (\ref{tan1}). However, in the other
limit, when $mR\gg 1$ the lowest eigenvalues are shifted down to
$(2n+1)\mu_0/2$. Then  the main component  of $\nu_L$ will have 
a mass of $\mu_0/2$ and the mixing angle will be in general 
$\tan\theta_n= N_0/N_n$.

In consequence, irrespective of the hierarchy of the masses in $M_D$, a
solution to the neutrino puzzles with bulk neutrinos  needs localization
of the three bulk neutrinos in different
branes with extra assumptions about the Yukawa couplings.

\section{Models with right handed neutrinos in the brane}

We now proceed to consider
the second class of models where we include both the
left and the right handed neutrinos in the same brane and  a blind bulk
neutrino which only couples with the right handed neutrinos due to either
gauge  symmetries~\cite{mnp} or  asymmetric boundary conditions 
in the bulk~\cite{mp}. 

The simplest gauge
model where this scenario is realized  is the left-right symmetric model
where the right handed symmetry is broken by the doublet Higgs bosons
$\chi_R(1,2,1)$, where the number inside the parenthesis correspond to the
quantum numbers under $SU(2)_L\times SU(2)_R \times U(1)_{B-L}$.
The relevant terms of the  action for one generation are 
 \be
 {\cal S}= \int d^4x [\kappa\bar{L}\chi_L \nu_B(y=0) +
 \kappa \bar{R}\chi_R
 \nu_B(y=0)+h \bar{L}\phi R] + \int d^4x dy
 \bar{\nu}_B\Gamma^5\partial_5\nu_B + h.c.
 \ee
As discussed in reference~\cite{mnp}, by setting  $<\chi^0_R>= v_R$ and
$<\chi^0_L>=0$, the following Dirac neutrino mixing  matrix is obtained
 \be
 (\bar{\nu}_{eL}~\bar{\nu}_{0BL}~  \bar{\nu}'_{BL})\left(\begin{array}{cc}
 hv & 0\\ \kappa v_R & 0 \\\sqrt{2}\kappa v_R &  \partial_5
 \end{array}\right)\left(\begin{array}{c}\nu_{eR} \\
 \nu'_{BR}\end{array}\right),
  \label{mright}
 \ee
with $\nu_{0BL}$ being the zero mode and $\nu'_{B}$ representing the
exited modes as before.
(Let us note parenthetically that in general a nonvanishing value for
$<\chi^0_L>$
could be expected from the potential, so we should assume either that 
this vev is small so that its contribution to neutrino masses is
negligible or
that the bulk neutrino breaks explicitly the parity symmetry so the
coupling with $L$ is zero as advocated in Ref.~\cite{mp}.)
Now, provided that $\kappa v_R\gg hv\simeq$ few MeV, 
a massless field which is predominantly the electron
neutrino appears. Since 
$\kappa\simeq \frac{M^*}{M_{P\ell}}$, this constraint implies that $M_*$ 
must be as large as $10^8$ GeV or so.  
Oscillations into bulk neutrino will now result, again 
dominated by the lowest
mass of the bulk modes~\cite{mnp}, implying that the largest radius of the
extra dimensions should still be at the millimeter range. 

Let us now look at the complete picture for three generations. The more
general mass terms in this class of models are
 \bea
 {\cal L} &=& \bar{\bf \nu}_L\cdot  {\cal M}_{LL} \cdot {\bf \nu}_{L}
 + \bar{\bf \nu}_L\cdot  M_D \cdot {\bf \nu}_{R}
 + \bar{\bf \nu}_R\cdot  M_N \cdot {\bf \nu}_{R} \nonumber \\
 & &  +  \bar{\bf \nu}_R\cdot  {\bf m} \cdot \nu_{BL}(y=0)
 + \int dy\, \bar{\bf \nu}_{BL}\cdot\partial_5 \cdot{\bf \nu}_{BR}
 + h.c.
 \label{3lr} 
 \eea
In last equation ${\cal M}_{LL}$ represent the Majorana mass terms of the
left handed
neutrinos. For simplicity we will assume those terms to be zero. $M_N$
represents the Majorana matrix for the right handed neutrinos which in the
left-right models arises from the vev of a $B-L=2$ $SU(2)_R$ triplet and 
${\bf m}^{\dagger}= m(1,1,1)$ are the universal couplings of the bulk
neutrinos
to the right handed neutrinos. 
The mass matrix now has the profile
 \be
 \left(\begin{array}{c c c c c}
  0          & M_D              &   0     &         0       & 0 \\
 M_D^\dagger & M_N              & {\bf m} & \sqrt{2}{\bf m} & 0 \\
  0          &  {\bf m}^\dagger &   0     &         0       & 0 \\
  0          &\sqrt{2}{\bf m}^\dagger & 0 &         0  & \partial_5 \\
  0          &    	0       &   0     &     \partial_5  & 0 
 \end{array}\right) 
  \label{m3}
 \ee
in the basis
$({\bf \nu}_L, {\bf \nu}_R,\nu_{0BL}, \nu'_{BL}, \nu'_{BR})$.
Notice that this is really an  infinite matrix with 
the entry $\sqrt{2}{\bf m}$ 
being the universal coupling
of $\nu_R$ to the exited bulk modes $\nu_{nBL}$, 
then, this is an infinite matrix itself given by 
$\sqrt{2}{\bf m}\times (1,1,\cdots)$ . 
We have used the two component notation and used subscripts $L,R$ as
labels for corresponding states rather than helicity projection operators. 
Notice that $\nu_{0BR}$ is massless and is therefore not included in the
above mass matrix. As a result, this matrix has odd number of rows and
columns and has one zero eigenvalue. The corresponding state is
identified due to its flavor content to be the close to the $\nu_e$
state. The elements of $M_N$ are expected
to be large, $\leq M_*$, while those in $M_D$ as well as $m$ are of the
order of some MeV. This via the usual seesaw mechanism leads to light
eigenstates.

We can write down the general mixing matrix for the ``four neutrino'' 
states as:
 \be 
 \left(\ba{c} \nu_e \\ \nu_\mu \\ \nu_\tau \\ \nu_B \ea \right)_L = 
 U \cdot 
 \left(\ba{c} \nu_1 \\ \nu_2 \\ \nu_3 \\ \nu_4 \ea \right)_L
 \label{four}
 \ee
where the neutrinos in the right hand side are the mass eigenmodes, and
$\nu_4$ represent the mass eigenstates of the KK tower. The active
neutrinos will contain in general suppressed  contributions from the
complete KK tower, then, only the lowest mode is expected to contribute
substancially and  we will at the end identify this mode with $\nu_4$.
We will show that in this picture,
  $|U_{e1}|\approx 1$ and $|U_{e4}|>|U_{e2,3}|$; $\nu_1$ being 
the lightest element of the spectrum with zero mass.  
If the masses of $\nu_{2,3}$ are
in the eV range with a mass difference square of the the order of $\Delta
m^2_{atm}$, we can solve the atmospheric neutrino puzzle if
$|U_{\mu 2}|/|U_{\mu 3}| \sim |U_{\tau 2}|/|U_{\tau 3}| \sim 1$, 
with the decoupling of the bulk modes from this sector; thus we would need 
$|U_{\mu,\tau 4}|\ll 1$ and $|U_{B4}|\approx 1$. 
Finally to explain the LSND data, we need 
$|U_{\mu 1}|\sim|U_{e2}|\gg|U_{e3,\mu 4}|$. We will show below that
all these features emerge only from the simple choice of the seesaw matrix
$M_N$ as is commonly done in most model building\cite{rev}.

Let us see how this scenario is realized in our picture. First
note that the mass matrix in (\ref{m3}) has a zero mass left handed eigen
mode, $\nu_{1L}$,  which is a linear combination of ${\bf \nu}_L$ and
$\nu_{0BL}$ given by the solution to the matrix equation
 \be
 \left(\ba{cc}M_D^\dagger ~ {\bf m}\ea\right)\cdot | \nu_{1L}> = 0.
 \ee
where $(\ba{cc}M_D^\dagger ~ {\bf m}\ea)$ is a $4\times 3$
matrix.
If we want this lightest eigenstate to be predominantly the electron
neutrino to maintain the observed universality in charged current weak
interactions, we must demand that in the expression for $\nu_{1L}$, 
\be 
 \nu_{1L} = {1\over N} \left[ \nu_{eL} + \delta \nu_{\mu L} 
 + \Delta \nu_{\tau L}  - \epsilon \nu_{0L}\right ]
 \label{nu0}
 \ee
we must have $\delta, \Delta, \epsilon \ll 1$. If we make 
the simplest choice for $M_D = Diag(m_1,m_2,m_3)$, with the natural
hierarchy $m_1<m_2<m_3$, we get $\delta = m_1/m_2 > \Delta=m_1/m_3$ and
$\epsilon = m_1/m$. We then may set $1>\epsilon >\delta\gg \Delta$ to get
$\sin^2 2\theta_{sol}\simeq \epsilon^2$, 
and $\delta$ is going to produce the mixing for the 
LSND data.

After the extraction of the zero
mass term (\ref{nu0}) we may procceed to decouple the right handed
fields which get large masses via $M_N$.
The mass matrix  for light fields then  reduces to
 \be
 \left( \ba {c c c}
    M_L M_N^{-1} M_L^\dagger   & \sqrt{2}M_L M_N^{-1}{\bf m}    &  0 \\
 \sqrt{2}{\bf m}^\dagger M_N^{-1}M_L^\dagger &  
  2{\bf m}^\dagger M_N^{-1}{\bf m} &  \partial_5 \\
    0  & \partial_5  &  0 
 \ea\right),
 \label{meff}
 \ee
in the  basis
$(\nu_{0BL}, \nu_{\mu L}, \nu_{\tau L}, \nu'_{BL}, \nu'_{BR})$;
with $M_L$ of the form
 \be
 M_L =  \left( \ba {c c c}
  m & m & m\\
  0 & m_2& 0\\
  0 & 0 & m_3 
  \ea \right)
  \ee
where we have neglected those contributions that are 
proportional to the small parameters $\epsilon$, 
$\delta$ and $\Delta$ to 
show how the decoupling of $\nu_{\mu,\tau}$ may occurs. 
It is worth noting that this form of
$M_L$ is produced by the conservation of $L$
in the Standard Model which suggest that $M_D$ should
be diagonal, thus giving $M_L$ as above. 
In this picture, the $L$ violation
terms come from high energy, and all the texture in the mixing matrix 
is produced through $M_N$. Also, notice that the mass matrix (\ref{meff})
is really of a generical form. It may be produced 
not only by see-saw as we are assuming here but as well as  through some
other mechanism as radiative corrections. 

Now, the main ingredient of the decoupling is based on the fact that the
bulk neutrino really couples to a certain linear combination of the brane
neutrinos, given by $M_L M_N^{-1}(\nu_{0B},\nu_\mu,\nu_\tau)_L^\dagger$,
and decouples from the  orthogonal combinations. Moreover, the first
row and column of $M_L M_N^{-1} M_L^\dagger$ correspond to the same
combination. Therefore by rotating the brane sector to 
diagonalize $M_L M_N^{-1} M_L^\dagger$, we may expect that the
decoupling becomes explicit. 
Lets see how this argument works. First notice that
if we make the rotation 
\be 
 \left(\ba{c} \nu'_{0B} \\ \nu_2 \\ \nu_3 \ea \right)_L = 
 K \left(\ba{c} \nu_{0B} \\ \nu_\mu \\ \nu_\tau \ea \right)_L,
 \ee
then  the mass matrix (\ref{meff}) transforms into 
 \be
 \left( \ba {c c c}
   K M_L M_N^{-1} M_L^\dagger K^\dagger   &
   	 \sqrt{2}~ K M_L M_N^{-1}{\bf m}    &  0 \\
 \sqrt{2}~{\bf m}^\dagger M_N^{-1}M_L^\dagger  K^\dagger &  
  2~{\bf m}^\dagger M_N^{-1}{\bf m} &  \partial_5 \\
    0  & \partial_5  &  0 
 \ea\right).
 \label{mk}
 \ee
Let us choose $K$ such that $K M_L M_N^{-1} M_L^\dagger K^\dagger$ becomes
diagonal. All the mixing with the bulk modes $\nu'_{BL,R}$ are then
given by the form of the column vector $K M_L M_N^{-1}{\bf m}$. 
It is here that we expect to see the decoupling.
Let us take an explicit form for $M_N$ to make it clear: 
 \be
 M_N = \left( \ba{c c c} 
 0 & a & 0 \\
 a & 0 & b \\
 0 & b & c \ea \right)
 \ee
and to simplify matters let us assume that $m=10$, $m_2=50$, $m3=100$ and
$a=10^{7}$, $b=5\times 10^{9}$ and $c=10^{8}$ all in MeV.  These values
are selected merely for illustrative purpose.
Other values may also achieve the desired decoupling. The
correct entries in the matrix depend on the high scale sector of the
model, which we leave unspecified for our purpose. 
Now we proceed with
the numerical computation  with above choice of numbers and
get 
 \be 
 K = \left(\ba{ccc}
    0.9998 & 2\times  10^{-4} & - 2\times  10^{-2} \\
    -1.4\cdot 10^{-2} & 0.7103 & -0.7038 \\
    - 1.4\cdot 10^{-2} &  -0.7039 & -0.7002
     \ea\right),
 \ee
while the eigenvalues of $M_L M_N^{-1} M_L^\dagger$ are  given by
(.25 MeV, -1.0028 eV, 1.0008 eV ).
Thus we get a maximal mixing in the right sector. 
On the other hand,  for
the mixing with the bulk modes we get
$K M_L M_N^{-1}{\bf m} = (.25, 1.4\cdot 10^{-8} , - 1.4\cdot 10^{-8})$
MeV. 
Thus, $\nu_{2,3}$ decouples while $\nu'_{0BL}$ mixes maximally to
the bulk modes. 

Next, 
as the entry ${\bf m}^\dagger M_N^{-1}{\bf m}$   in (\ref{meff})  also
is the first element in the mixing $M_L M_N^{-1}{\bf m}$, 
after extracting the decoupling modes, the effective mass
matrix  reduces to the profile
\be
 \left( \ba {c c c}
    \alpha   & \sqrt{2}\alpha    &  0 \\
 \sqrt{2}\alpha  &  
  2\alpha  &  \partial_5 \\
    0  & \partial_5  &  0 
 \ea\right) \equiv
 \left( \ba {c c c c c c}
    \alpha   & \sqrt{2}\alpha    &  0  & \sqrt{2}\alpha    &  0 &\cdots\\
 \sqrt{2}\alpha  &   2\alpha  & \mu_0 & 2\alpha  &  0 & \cdots \\
    0  & \mu_0  &  0  & 0 & 0 & \cdots \\
    \sqrt{2}\alpha  &   2\alpha  & 0 & 2\alpha  & 2 \mu_0 &\cdots \\
    0  & 0  &  0  & 2 \mu_0 &   0 & \cdots\\
    \vdots & \vdots &\vdots & \vdots &\vdots &\ddots
 \ea\right)
 \label{mheavy}
 \ee
where  $\alpha = {\bf m}^\dagger M_N^{-1}{\bf m}$.
It is simple to check this fact in  the case above since
$\alpha \approx .25$ MeV too. It is worth noting that now all the fields
involved are bulk modes.
The rows and columns in the left hand side   of (\ref{mheavy}) are   
labeled by $(\nu_{0L}, \nu'_{BL}, \nu'_{BR})$. 
On the right hand side  we 
have explicitly written it in its true infinite mass matrix form labeled
by $(\nu_{0L}, \nu_{1BL}, \nu_{1BR}, \nu_{2BL}, \nu_{2BR},\cdots)$. 

We may worry about the high value of $\alpha$ since it could mean that
$\nu_{0BL}$ has a component with a large mass; however, as a result of
the  maximal mixing with the KK modes, that contribution  will be
suppressed as well as the other heavy elements of the tower are. 
The mechanism
works exactly as in the one generational case
of this class of
models~\cite{mnp}  where the $\nu_L$ which gets mixed with the bulk modes
has a mass term in the MeV range [see Eq. (\ref{mright})]. 
There, the heavy term is absorbed by the tower and as a result 
the masses of the KK modes are shifted down, 
while their contributions to $\nu_L$ becomes
suppressed as $1/n$. This  suppresses the 
heavy field contribution by the number of levels
below such a mass i.e. $\sim \alpha R$ MeV, which is a large number. 
The lowest mode then 
becomes the main component of $\nu_{0L}$ that appears in Eq. (21);
We identify this mode as the sterile neutrino, $\nu_4$,
used in  solving the solar neutrino puzzle.


Let us demonstrate this by 
diagonalizing the mass matrix (\ref{mheavy}). We take advantage
of this compact notation to write the characteristic equation as
 \be
 (\lambda^2 - \partial_5^2)\left[ \alpha - \lambda + 
 	{2\alpha\lambda^2 \over \lambda^2 - \partial_5^2} \right] = 0
 \ee
which now reduce to $\pi\alpha R\cot(\pi\lambda R) = 1$
and may be solved exactly to get the mass eigenvalues
 \be 
 \lambda_n = {{\rm \arctan} (\pi\alpha R)\over \pi R} + 
 {n\over R}.
 \ee
where now $n \in {\cal Z}$, since also the right handed bulk fields 
are involved.
Clearly, when $\alpha R\gg 1$ (as is the case here) the masses are just 
$\lambda_n = (2n+1)\mu_0/2$, which are now shifted down by 
${1\over 2} \mu_0$. Also we note that 
the only masses of order of $\alpha$ correspond to 
$n\approx \pm \alpha R$ which were already in the tower; thus our massive 
term has been totally absorbed into the KK tower. Strictly speaking, there
 is an extra mode in the tower with infinite mass. 

The corresponding eigenvectors are given
by ($\lambda$ are the eigenvalues)
 \be
 \nu (\lambda) = {1\over\eta(\lambda)}\left[\nu_{0BL} + 
  {\sqrt{2}\lambda^2\over \lambda^2 - \partial_5^2} ~ \nu'_{BL}
  + {\sqrt{2}\lambda \partial_5 \over \lambda^2 - \partial_5^2}~\nu'_{BR}
  \right].
 \ee
with the normalization factor: 
 \be
 \eta^2 = \left({\lambda\over \alpha}\right)^2 
 \left[ 1 + (\pi\alpha R)^2\right] 
 \approx (2n+1)^2 \left({\pi\over 2 } \right)^2.
 \label{eta}
 \ee
Notice that  in our limit i.e. $\alpha R\gg 1$,
the masses are degenerate, and occur in pairs with 
masses $\pm \lambda_n$. We may then recombine those states to get two
Majorana neutrinos that form a Dirac neutrino of mass $\lambda_n$ of the
form
 \be
 \tilde\nu_{nL} = 
 {\sqrt{2}\over\eta(\lambda_n)} \left[\nu_{0BL} + 
  {\sqrt{2}\lambda_n^2\over \lambda_n^2 - \partial_5^2} ~ \nu'_{BL}
  \right]; 
  \qquad 
  \tilde\nu_{nR} = {1\over\eta(\lambda_n)}~
  {2\lambda_n \partial_5 \over \lambda_n^2 - \partial_5^2}~\nu'_{BR}
 \ee
Now, we may
rotate the system backwards to write down $\nu_{0BL}$ in terms
of the left handed mass eigenstates  to get
 \be 
 \nu_{0BL} = 
 \sum_{n=0}^\infty {\sqrt{2}\over \eta(\lambda_n)} ~\tilde \nu_{nL} .
 \ee
Thus, as $\eta$ grows quadratically in $\lambda_n$, 
the main element of
$\nu_{0BL}$ is the lowest mode $\tilde\nu_{0L}$ that we then 
identify as the  sterile neutrino. 
Moreover, from (\ref{eta}) we see that the 
suppression over the other modes goes as $1/(2n + 1)$, then the heavier
modes get highly suppressed, since $n$ is also the number of eigenlevels
below.
 
Finally, we may put all the ingredients together and calculate the 
effective mixing pattern at low energies. We have proceeded using a
numerical analysis from
the beginning to estimate masses and mixings directly from the mass matrix
in (\ref{m3}), without the approximations we made in the former 
analysis to
get more accurate values. 
The input values for our parameters are the same as before
with $m_1=1$ MeV 
and we have taken $\mu_0 = 5\times 10^{-3}$ eV to fit properly
$\Delta m^2_{sol}$ ,  we then get
 \be
 \left(\ba{c} \nu_e \\ \nu_\mu \ea \right) = 
  \left(\ba{c c c c } 
  0.9948 & 0.0212 & -0.0072 & 0.0632  \\
  0.0199 & -0.7037 & 0.7102  & 0.0013  \ea \right)  
 \left(\ba{l} \nu_1( {\rm massless} ) \\ \nu_2 (-1.0014~eV) \\
              \nu_3(1.003~eV) \\ \nu_4( 2.5151\cdot 10^{-3}~eV)
   \ea \right) 
 \ee
where the masses are indicated between brackets. From these results we get
 \be
 \ba{l l c l l} 
 \Delta m^2_{sol}&\sim 6.326 \times 10^{-6}~ eV^2, &\qquad &
 \sin^2 2\theta_{sol} &\sim 1.58 \times 10^{-2}  \\
 \Delta m^2_{atm}&\sim 3.23 \times 10^{-3}~ eV^2, & &
 \sin^22\theta_{atm} & \sim .9991\\ \nonumber
 \Delta m^2_{e\mu}&\sim 1.0028~ eV^2, \quad & &
 \sin^22\theta_{e\mu} & \sim1.18\times 10^{-3}
 \ea
 \ee 
These values are to be compared with the experimentally preferred values
(\ref{sol}), (\ref{atm}) and (\ref{emu}).  It clearly shows that the
choice of the bulk neutrino as the sterile neutrino works quite well to
explain observations. Although for the sake of explicit demonstration we
have made a specific choice of parameters, there is a range of parameters
in the mass matrix that will do equally well.

\section{Conclusions}

In conclusion, our analysis seems to demonstrate that one can use the bulk
neutrino as a sterile neutrino needed to understand all neutrino
oscillation observations. Despite the fact that it has an infinite set of
KK excitations, only the lowest mode seems to play an important role and
indeed gets identified as the sterile neutrino.
It is worth emphasizing again the conceptual advantage of this approach
which is that the ultralightness of the sterile neutrino has a geometric
origin rather than from extra symmetries, as in the four dimensional
models. What is also gratifying is that the apparent connection between 
submillimeter gravity experiments and the neutrino puzzles seems to remain
in the realistic implementations of this idea.

\vskip1em

{\it Acknowledgements.}  
The work of RNM is supported by a grant from the National
Science Foundation under grant number PHY-9802551. 
The work of APL is supported in part by CONACyT (M\'exico).


\end{document}